# Are all Quasi-static Processes Reversible?

**Debasis Mukhopadhyay[1], and Kamal Bhattacharyya [2],***

[1]Department of Chemistry, University of Calcutta, Kolkata – 700009, India
E-mail: dmukhopadhyay2000@yahoo.com

[2]Department of Chemistry, University of Calcutta, Kolkata – 700009, India
E-mail: pchemkb@yahoo.com

**Abstract:** A process, carried out in a stepwise manner, becomes quasi-static when the number of intermediate steps tends to infinity. Usually, the net entropy production approaches zero under this limiting condition. Hence, such cases are termed reversible. A favorite example is the introduction of an infinite number of intermediate-temperature reservoirs in between the source and the sink for a non-isothermal heat transfer process. We analyze the situation and conclude that such quasi-static processes are not reversible. Indeed, no non-isothermal heat transfer process can ever be made reversible due to an extraneous work term.

*Author for correspondence

**I. INTRODUCTION**

In thermodynamics, characterizing the reversibility of a process continues to be of interest.[1-10] The concept of a quasi-static process (QSP) is less critically introduced and classic works[1-4] do not shed sufficient light to distinguish a QSP from a truly reversible process (RP). Several kinds of misconceptions arise from this lack of clarity. Here, we particularly concentrate on the question: How should one distinguish an RP from a QSP?

A QSP proceeds 'through a dense succession of equilibrium states'[3]. Such a view is generally adopted in other works[1,4] as well. These processes occur *infinitely slowly* and, according to Max Born, as quoted in Ref. 6, 'one should call such processes quasi-static, but one usually uses the word reversible, as they normally have the property of being reversible.' The notion prevalent among us is clear from this statement. Born, however, bypassed what should we mean by 'normally', and when abnormal situations would crop up. Caratheodory (see Ref. 6) stressed the role of *external work*, or equilibrium forces, during a QSP. He also remarked on QSP in presence of internal friction. This latter part was elaborated subsequently[5] and distinguished from an RP. It is now accepted that a QSP is not equivalent to an RP in presence of dissipation.[4] On the other hand, an RP is defined as a process that can 'take place in the reverse direction'.[2] But, this does not seem to be a complete statement. Indeed, we should strengthen this definition by additionally requiring that the magnitudes of heat and work terms should turn out to be exactly same in the reverse process, only with their signs reversed. Callen[3] unified RP and QSP by remarking that 'the limiting case of a quasi-static process in which the increase of entropy becomes vanishingly small is called a reversible process'. This view seems to be widespread in the literature[7-10] these days.

Let us first quickly clarify a few points before concentrating on the main issue. It is true that any RP is also a QSP. But, *all* such processes do *not* require infinite time for completion. Phase changes are natural examples that are reversible, and occur within finite times. Furthermore, we should be cautious in concluding that all QSP become equivalent to RP under zero entropy production limits. Indeed, there exists a natural requirement for an RP. This is to couple the process with a few other RP to form an engine, and then to test if the heat and work terms involved there are *exactly* reversed when we choose the device to act as a pump. What is more, for an irreversible process coupled in the above manner, a gradual increase in efficiency of the engine will result as the number of steps increases. Most interesting irreversible process in this regard would be an *isochoric* one for a gas where mechanical *work* term, which was central in Caratheodory's statement, is absent. We shall, therefore, consider the performance of some such engines as well. Additionally, one may be curious to compare how the efficiency of one such engine is related to the Carnot efficiency. This is particularly relevant[11,12] because all theoretical discussions involving heat engines start with the Carnot engine (CE). It is an idealized heat engine that works reversibly between a high temperature ($T_H$) reservoir (source) and a low-temperature ($T_L$) one (sink), with an efficiency $\eta_{CE}$ independent of the working substance. This result is well established. However, it is important to recall what happens when we consider one or more steps of an engine to occur irreversibly. We know that any such engine will have efficiency $\eta < \eta_{CE}$ and there will be some net entropy production. Further, an excellent analysis[13] reveals that there is no direct relationship between $(\eta_{CE} - \eta)$ and the net entropy production, as long as both are non-zero. But, when one insists the irreversible steps to occur as QSP,[7-9] associated with zero entropy production, one would seem to infer that, under such a condition, $\eta$ would become equal to $\eta_{CE}$ for the engine at hand. The standard technique to achieve this end is to incorporate a large number (theoretically infinite) of intermediate-temperature reservoirs between $T_H$ and $T_L$.[7,9] For example, the effect of insertion of an infinite number of reservoirs either in arithmetic progression or in geometric progression for a cooling or heating process has been noted.[9] Particular reference may be made in this context to the Otto cycle[4] and the Stirling engine (SE),[11,12] both of which involve irreversible isochoric steps. In these works, the chief idea has been to establish that non-isothermal heat transfer processes can be made reversible. It is precisely this point that we like to scrutinize. A related concept in this context is to use regenerators.[12,14] They improve the efficiency of certain engines, but are more of technical interest. Therefore, we shall briefly consider their role.

Our organization is as follows. In section II, we have scrutinized the effect of insertion of a large number of intermediate temperature reservoirs on the entropy production, and then on the heat as well as work involved, and hence the efficiency, for two different engines containing non-isothermal heat transfer steps, along with the role of regenerators in this context. Section III is reserved for the analysis of reversibility criterion. Here, we finally trace the error involved in the simplistic 'zero entropy production limit' argument. We summarize our findings in section IV. Two sample problems are provided at the end in support of our analysis.

## II. QUASI-STATIC PROCESS UNDER A LIMIT

We consider here two processes. The first is a simple heating process of $n$ moles of a solid substance of heat capacity $C$ from $T_1$ to $T_2$ ($T_1 < T_2$). The second is an isochoric process of a gas that may involve either heating or cooling, but is associated with some other reversible processes to form cycles. We choose two such cycles. In each case, the whole set-up acts as an engine. In these two engines, our choice of the system will be $n$ moles of an ideal gas with $C_V = \alpha R$ where '$\alpha$' is a positive constant.

### A. Simple heating

Let us take the heating case. We have the following results for $\Delta S$ of the system and surroundings (here the reservoir at $T_2$):

$$\Delta S_{sys} = nC \ln \frac{T_2}{T_1}, \tag{1}$$

$$\Delta S_{sur} = -nC \frac{T_2 - T_1}{T_2}. \tag{2}$$

Calling $(1 - T_1/T_2)$ as $x$, we have

$$\Delta S_{tot} = -nC [\ln(1-x) + x] > 0. \tag{3}$$

The process is therefore irreversible.

An additional reservoir at $T$, ($T_1 < T < T_2$), would not change Eq. (1). But, taking part in this process, it can change Eq. (2). We now have

$$\Delta S_{sur}^1 = -nC\frac{T_2 - T}{T_2} - nC\frac{T - T_1}{T}. \tag{4}$$

One can check that $\left|\Delta S_{sur}^1\right| > \left|\Delta S_{sur}^0\right|$ where the superscript signifies the number of intermediate-temperature reservoirs. The net effect on $\Delta S_{tot}$ may be summarized as

$$\Delta S_{tot}^0 > \Delta S_{tot}^1 > 0. \tag{5}$$

Thus, the net entropy production gets reduced. Here, $\Delta S_{tot}^0$ refers to Eq. (3).

Suppose, we now insert ($N - 1$) number of heat reservoirs in between $T_1$ and $T_2$ at equal temperature gaps of $\varepsilon$. Therefore, we have, $T_2 = T_1 + N\varepsilon$. The $i^{th}$ intermediate reservoir is at temperature $(T_1 + i\varepsilon)$. The entropy change $\Delta S_{sur}$ for the $i^{th}$ heat reservoir is given by

$$(T_1 + i\varepsilon)\Delta S_{sur} = -nC\varepsilon. \tag{6}$$

Hence, the total change in entropy of the surroundings would be obtained as

$$\Delta S_{sur}^{N-1} = -nC\sum_{i=1}^{N}\frac{\varepsilon}{T_1 + i\varepsilon}. \tag{7}$$

The sum in Eq. (7) can be estimated for large $N$ by taking the leading correction of Euler-Mclaurin sum formula.[15] This gives the result in the limit $N \to \infty$

$$\Delta S_{sur}^{N-1} \cong -nC\ln\frac{T_2}{T_1} + \frac{nC\varepsilon}{2}\left(\frac{1}{T_1} - \frac{1}{T_2}\right). \tag{8}$$

Coupling with Eq. (1), we get

$$\Delta S_{tot}^{N-1} = \frac{nC\varepsilon(T_2 - T_1)}{2T_1T_2}. \tag{9}$$

In the limit $N \to \infty$, one can take $\varepsilon \to 0$ because $N\varepsilon$ is finite. Thus, Eq. (9) shows that we can reduce the net entropy production to zero. The process therefore becomes a QSP. Apparently, this is in tune with the standard assumption that a sufficiently slow change of external parameters would be associated with zero entropy production. We should, however, point out that there exist cases[16,17] where one can perceive phenomena in contradiction to this maxim. Later, this point will concern us again. Important also is the role of time[18,19] that is required to achieve equilibrium at each point. It is because of this time lag that the process requires an infinite time. Incidentally, it has been pointed out[18] that some specific RP may not qualify to be a QSP, but in a physically uninteresting situation.

**B. Isochoric processes**

First, we consider a cycle with two isochoric processes. This is the SE consisting of four total steps:

I. An isothermal expansion A → B: $[P_1, V_1, T_1] \to [P_2, V_2, T_1]$
II. An isochoric cooling B → C: $[P_2, V_2, T_1] \to [P_3, V_2, T_2]$
III. An isothermal compression C→ D: $[P_3, V_2, T_2] \to [P_4, V_1, T_2]$
IV. An isochoric heating D → A: $[P_4, V_1, T_2] \to [P_1, V_1, T_1]$

For reversible isothermal heat transfer steps I and III, as taken in the SE, $\Delta S_{tot} = 0$. The other two steps need attention. Here, $T_1 > T_2$. In figure 1, ABCDA refers to this cycle.

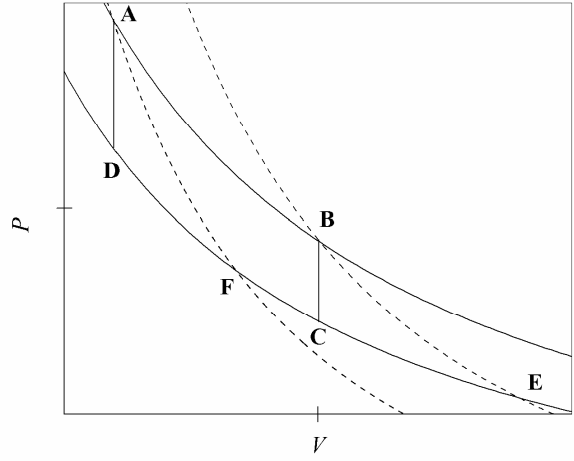

*Figure 1.* A pressure-volume curve of an ideal gas showing ABCDA as a Stirling cycle and CBEC as another three-step cycle. The dotted lines indicate adiabatic processes.

Consider the isochoric cooling step (step II) of the SE. Here, $\Delta S$ of the system is

$$\Delta S_{sys} = -nR\alpha \ln\left(1 + \frac{T_1 - T_2}{T_2}\right) \tag{10}$$

implying a loss of its entropy. The sink, in turn, has a positive $\Delta S$, given by

$$\Delta S_{sur} = nR\alpha(T_1 - T_2)/T_2 \tag{11}$$

and thus the net change is

$$\Delta S_{tot} = nR\alpha\,[x - \ln(1+x)] > 0 \tag{12}$$

where, we now define, $x = (T_1 - T_2)/T_2$.

For the isochoric heating step (step IV) of the SE, we obtain for the system

$$\Delta S_{sys} = nR\alpha \ln\left(1 + \frac{T_1 - T_2}{T_2}\right) \tag{13}$$

and for the source

$$\Delta S_{sur} = -nR\alpha(T_1 - T_2)/T_1 \quad (14)$$

so that the total change becomes

$$\Delta S_{tot} = -nR\alpha\,[\,y + \ln(1-y)\,] > 0;\ y = x/(1+x). \quad (15)$$

If we now consider the engine as a whole, the net $\Delta S$ would take account of changes of the reservoirs only. This is given by

$$\Delta S_{tot} = nR\alpha\,(x - y) = nR\alpha\,x^2/(1+x) > 0. \quad (16)$$

Let us note that, by introducing extra reservoirs, as demonstrated in Sec. II.A., one can reduce the net entropy production in Eq. (16). In the limiting situation of a QSP, Eq. (16) would tend towards zero as well. But, such an arrangement does *not* change the basic characteristics (heat and work terms) of the *engine*. The engine still takes up $Q_1$ heat in step I and does $W_1$ amount of work, rejects $Q_2$ heat in step II and $Q_3$ in step III on doing $W_3$ work on it, and finally takes up $Q_4$ heat in step IV. The efficiency of a heat engine is defined as the fraction of the total heat absorbed in the cycle that is converted into *net* work done per cycle. Therefore, efficiency of the SE is given by

$$\eta_{SE} = \frac{(Q_1 + Q_4) - (Q_2 + Q_3)}{Q_1 + Q_4}, \quad (17)$$

and it does *not* change by introducing extra reservoirs between $T_1$ and $T_2$. Further, standard relationships for the dashed lines (reversible adiabatics) of the figure yield $V_F/V_E = V_A/V_B = V_1/V_2$. Thus, $Q_3$ (SE) = $Q_3$ (CE). Such equality also holds for $Q_1$. We also have $Q_2 = Q_4$. It is hence simple to show that

$$\eta_{SE} < \eta_{CE} = \frac{Q_1 - Q_3}{Q_1}. \quad (18)$$

Primarily, relation in Eq. (18) holds[11,12] because, in isochoric steps of the SE (in which it differs from the CE), by choice, heat transferred is a path-independent function and so remains unchanged even if additional reservoirs are introduced at intermediate temperatures along the isochoric path. For such insertions, the area on the *P-V* graph (see figure 1), signifying the net work, also remains unchanged and thus the value of $\eta_{SE}$ remains unchanged, in spite of the decrease in entropy production. This simple qualitative argument also explains that the decrease in entropy production may *not* lead to an increase in efficiency of an irreversible engine,[13] *even* in the QSP limit.

We secondly consider a single isochoric process. Indeed, Figure 1 renders such a possibility along the cyclic path CBEC. This is similar to cycle DAFD and hence one would suffice for future consideration. We shall stick to the former one. This engine consists of the following processes:

I. An isochoric heating C → B: $\qquad [P_3, V_2, T_2] \to [P_2, V_2, T_1]$

II. An adiabatic expansion B → E: $\quad [P_2, V_2, T_1] \rightarrow [P_E, V_E, T_2]$

III. An isothermal compression E→ C: $\quad [P_E, V_E, T_2] \rightarrow [P_3, V_2, T_2]$

Suppose the first step requires $Q_{CB}$ heat, the system does $W_{BE}$ work at step II and work $W_{EC}$ is done on the system in step III to reject heat $Q_{EC}$ at this step. It is elementary to check that the efficiency of this engine will be

$$\eta = 1 - \frac{T_2}{T_1 - T_2} \ln \frac{T_1}{T_2}. \tag{19}$$

Introducing variables $x$ and $y$, respectively defined by $x = T_2/T_1$ and $y = 1 - x$, and using the inequality $\ln(1 - y) < -y$, one can show that $\eta$ in (19) satisfies $\eta < \eta_{CE}$. Along CB, here too we can place extra reservoirs to reduce the net entropy production and that may vanish in the quasi-static limit, but we notice that the heat and work terms involved in this engine do not change and hence its $\eta$ cannot increase.

## C. Role of regenerators

Regenerators are add-on devices to engines or pumps.[12,14] In the case of the SE, for example, one may use a lump of metal mesh[12] that takes up heat during path BC so that the sink need not have to take any heat to increase its entropy. Again, this heat, reserved in the regenerator, is used during the path DA. Thus, entropy of the source does not increase. In this way, one avoids the net entropy production. To state otherwise, we now have $Q_2 = 0 = Q_4$ in Eq. (17) and hence $\eta_{SE} = \eta_{CE}$ follows. However, here we should be cautious in defining our system. Originally, we had the ideal gas as the system. Now, when we insist that $Q_2 = 0 = Q_4$, we mean to imply that our system is now the ideal gas *plus* the regenerator. Then, of course, paths BC and DA would be adiabatic.[12] But, note that the regenerator is disconnected during path CD. Otherwise, it must have given all its stored heat to the sink. Therefore, technically we achieve the Carnot efficiency by employing a SE with regenerator, only at the cost of defining the *working substance* properly. Thus we go outside the realm of thermodynamics. This point seems to be overlooked in discourses dealing with regenerators. What we can finally say is that, a regenerator can improve the efficiency of *certain* engines (see below), but it is a technological trick. It has nothing to do with the basic theorems and corollaries of thermodynamics.

We close this subsection by noticing that the idea of using a regenerator to increase $\eta$ of an irreversible engine does not work *universally*. Our three-step cycle CBEC is a clear case in point where there is just *one* isochoric heating step and no cooling one of the same sort.

## III. THE REVERSIBILITY CRITERION

A thermodynamic process involves both the system and its surroundings. Hence, it is surprising to demand that non-isothermal heat transfer processes considered above could be made reversible by changing the surroundings alone. We shall pay attention to this specific point immediately below. Next, we seek a better criterion of reversibility that makes it distinctive from a QSP. We shall then scrutinize the heat conduction problem of Sec. II.A more closely.

**A. The zero entropy production limit**

A simple heating process was discussed in Sec. II.A. This is present also in Sec. II.B as part DA of the SE and in Sec. II.C as part CB of the three-step cycle. In all these cases, the heat taken up by the system, and hence $\Delta S_{sys}$, is fixed. On the way to make it a QSP, we have added an infinite number of reservoirs with gradually increasing temperatures so as to increase the magnitude of $\Delta S_{sur}$ that itself is negative. In other words, we have here an infinite number of sources, instead of a single one, spanning a range of temperatures between $T_1$ and $T_2$. For a cooling process, a similar arrangement lowers positive $\Delta S_{sur}$ by utilizing infinite number of sinks. Particularly for the SE, we note that the arrangement blurs the distinction between the source and the sink. Therefore, the zero entropy production limit is achieved by (i) keeping the system's changes undisturbed and (ii) blurring the distinction between the source and the sink. The point is, whether this limit allows a QSP to become equivalent to an RP.

**B. Engines and pumps**

We stated before that a hallmark of an RP is the *exact* reversal in sign of the heat and work terms involved. Particularly for an engine, it is easy to check. If an engine draws $Q$ heat to perform $W$ amount of work and rejects the rest, and if *all* the steps comprising the engine are *reversible*, then the same engine can be run backwards as a pump with the *same* 'efficiency', i.e., it should be able to reject $Q$ heat by absorbing $(Q - W)$ heat when $W$ work is done on the device. Let us check what happens to the above clause when we consider the cycles ABCDA and CBEC. Table 1 shows the detail. Here, the fourth column gives the work done by the engine or the work to be done on the pump, as the case may be. The third row in each cycle [denoted by Pump*] shows entries that would have appeared had the original engines been *reversible*. Clearly, none is reversible.

*Table 1. Performance of the cycles (see Sec. II B) as engines and pumps*

| Cycle | Device | Heat absorbed | Work done | Heat rejected |
|-------|--------|---------------|-----------|---------------|
| ABCDA | Engine | $Q_1 + Q_4$ | $W_1 - W_3$ | $Q_2 + Q_3$ |
| DCBAD | Pump | $Q_3 - Q_4$ | $W_1 - W_3$ | $Q_1 - Q_2$ |
|       | Pump* | $Q_2 + Q_3$ | $W_1 - W_3$ | $Q_1 + Q_4$ |
| CBEC  | Engine | $Q_{CB}$ | $W_{BE} - W_{EC}$ | $Q_{EC}$ |
| CEBC  | Pump | $Q_{EC} - Q_{CB}$ | $W_{BE} - W_{EC}$ | 0 |
|       | Pump* | $Q_{EC}$ | $W_{BE} - W_{EC}$ | $Q_{CB}$ |

One may argue, on the basis of the above chart, that the SE with QSP along BC and DA works actually reversibly, because it then virtually takes no heat from the *highest* temperature reservoir and rejects nothing to the *lowest* temperature reservoir. Therefore, both $Q_2$ and $Q_4$ are zero. The second and third columns thus agree. We also then have $\eta_{SE} = \eta_{CE}$. But, this result is not acceptable on two grounds. First, in presence of many reservoirs, the Carnot efficiency should not follow.[13] Secondly, we cannot argue likewise that $Q_{CB} = 0$ for the second engine. One can see from the table that such a choice would violate the second law directly in course of its action either as an engine or as a pump.

Moreover, reversibility should not be based on processes occurring just *at* the highest and lowest temperature reservoirs. Indeed, a more convincing argument now follows.

## C. Extraneous work in non-isothermal heat transfer

A careful consideration of the heat transfer process reveals that mere addition of extra reservoirs is not enough to approach a QSP. Let us take the heating case DA in the SE (any other situation would be similar). The system at point D is connected to reservoir at $T_2$ maintaining thermal equilibrium. Its transition to point A is achieved by switching off the connection to reservoir at $T_2$ and switching on the connection with reservoir at $T_1$. The work done in operating *two* switches, say *w*, is *negligible* compared to the net work by the engine. With one extra reservoir, *four* switching operations are needed. But, as more and more intermediate-temperature reservoirs are introduced, increasing number of switching operations is to be performed and, in the QSP limit ($N \to \infty$), this would definitely cost a sizeable amount of work of magnitude *Nw*. What is more, when the process is reversed (*i.e.*, cooling), the same amount of work is again to be *done* from outside. The work done earlier is *not* retrieved. In this sense, there is an inherent asymmetry in these non-isothermal heating or cooling processes in presence of infinite reservoirs. Therefore, a QSP in forward direction costs some work. The reverse QSP also costs the same amount of work. Here lies a hidden irreversibility and this fact precisely makes such QSP different from a true RP.

## IV. CONCLUSION

To summarize, we have found that a QSP becomes equivalent to an RP not just when the net entropy production is zero. This is one of the necessary conditions. The other, and probably more general, is to couple the process with requisite number of reversible isothermals and adiabatics to form a cycle, and then to check if the work and heat terms of the cycle remains same in magnitude during its action as an engine and a pump. Indeed, when this condition is satisfied, the first one is automatically taken into account. Two more points need to be stressed in this context. First, if the number of steps of an irreversible process increase, either both the heat and work terms, or, at least, the work term, involved in the cycle should change, leading to an increase in efficiency of the whole cycle. In the QSP limit, this should attain the maximum value as well. Second, any non-isothermal heat transfer process is inherently irreversible in character. The first point becomes obvious when we deliberately choose a CE with any one branch irreversible. When an isothermal part is taken to be irreversible and is then performed in a stepwise manner, both heat and work terms change. As we increase the number of intermediate steps, the changes continue to raise the efficiency. Similar is the situation with an adiabatic branch, except that here *only* the work term would be influenced. The second point has its origin at the work involved in the on-off switches connected to an infinite number of heat reservoirs with which the system has to thermally equilibrate at different instants. This work term is outside the scope of thermodynamics, very much like the case of regenerators discussed before. In fine, we notice the importance of the *work* term, as Caratheodory stressed, in the context of a QSP. A QSP has the prime advantage of *tracing* the full path of a process only.

**SUPPLEMENTARY PROBLEMS:**

The following two problems are meant to demonstrate the inherently irreversible character of non-isothermal heat transfer processes even when the temperatures of two reservoirs, and hence the states involved, differ infinitesimally. Use ideal gas as the working substance in both the cases.

1. Consider the cycle CBEC of Fig. 1 and choose $T_2 = T_1 - \varepsilon$. Then, show that the efficiency of this engine is half the same of the Carnot engine when $\varepsilon \to 0$. Interpret the result.

2. Choose a cycle with one isothermal, one adiabatic and one isobaric step. As above, consider the cycle to work between two temperatures differing infinitesimally. Show again that the efficiency of this engine is half the same of the Carnot engine.